# Observation of Helical Edge States and Fractional Quantum Hall Effect in a Graphene Electron-hole Bilayer

J. D. Sanchez-Yamagishi[1]†, J. Y. Luo[1]†, A. F. Young[2], B. Hunt[3], K. Watanabe[4], T. Taniguchi[4], R. C. Ashoori[1], P. Jarillo-Herrero[1]*

[1]Department of Physics, Massachusetts Institute of Technology, Cambridge, MA 02139.
[2]Department of Physics, University of California Santa Barbara, CA 93106.
[3]Department of Physics, Carnegie Mellon University, Pittsburg, PA 15213.
[4]Advanced Materials Laboratory, National Institute for Materials Science, Tsukuba, Ibaraki 305-0044, Japan.

*Corresponding author. E-mail: pjarillo@mit.edu
†These authors contributed equally to this work.

## Abstract

A quantum Hall edge state provides a rich foundation to study electrons in 1-dimension (1d) but is limited to chiral propagation along a single direction. Here, we demonstrate a versatile platform to realize new 1d systems made by combining quantum Hall edge states of opposite chiralities in a graphene electron-hole bilayer. Using this approach, we engineer helical 1d edge conductors where the counterpropagating modes are localized in separate electron and hole layers by a tunable electric field. These helical conductors exhibit strong nonlocal transport signals and suppressed backscattering due to the opposite spin polarizations of the counterpropagating modes. Moreover, we investigate these electron-hole bilayers in the fractional quantum Hall regime, where we observe conduction through fractional and integer edge states of opposite chiralities, paving the way towards the realization of 1d helical systems with fractional quantum statistics.

A helical 1d conductor is an unusual electronic system where forward and backward moving electrons have opposite spin polarizations. Theoretically, a helical state can be realized by combining two quantum Hall edge states with opposite chiralities and opposite spin polarizations(*1, 2*). Most experimental efforts though have focused on materials with strong spin-orbit coupling(*3-5*), which avoids the need for magnetic fields. However, an approach based on quantum Hall edge states offers greater flexibility in system design with less dependence on material parameters. Moreover, a quantum Hall platform could harness the unique statistics of fractional quantum Hall states. Recent proposals have predicted that such a system, in the form of a fractional quantum spin Hall state(*6-8*), could host fractional generalizations of Majorana bound states.

To simultaneously realize two quantum Hall states with opposite chiralities, it is necessary to have coexisting electron-like and hole-like bands. Such electron-hole quantum Hall states are observed in semi-metals but suffer from low hole-mobilities (*9, 10*). In this respect, graphene is an attractive system because it has high carrier mobilities and is electron-hole symmetric. In fact, the graphene electron and hole bands can be inverted by the Zeeman effect to realize helical states (*11, 12*), but requires very large magnetic fields (*13, 14*). A similar outcome could be realized more easily in a bilayer system, where an electric field can dope one layer into the electron band and the other into the hole band. In a moderate magnetic field, this electron-hole bilayer will develop quantum Hall edge states with opposite chiralities in each layer. Here, we demonstrate a graphene electron-hole bilayer, which we use to realize a helical 1-dimensional conductor made from quantum Hall edge states.

All studied devices consist of two monolayer graphene flakes stacked together using a dry transfer process (*15, 16*). The stacking results in a rotational misalignment, or twist, between the layers. The dominant effect of the twist is to decouple the layers by separating their Fermi surfaces in momentum space (*17, 18*) (see Figure 1A). For twist angles larger than a few degrees, the band structure at low energies is given by two sets of Dirac cone dispersions which are each localized on a different layer despite the small *0.34* nm interlayer spacing. The layer decoupling persists in a magnetic field, with each layer developing a Landau level spectrum similar to monolayer graphene (*18-21*).

We fabricated a dual-gated structure where the twisted bilayer is encapsulated by hexagonal boron nitride (hBN) dielectric layers (Figure 1B). The hBN also shields the graphene layers from contamination during the fabrication process, resulting in a more homogenous electronic system (*15, 22*). The twisted bilayer is contacted using graphite electrodes, which makes good contact to both layers of the bilayer, even at high magnetic fields (further details in Supplementary Materials). Unless otherwise noted, all measurements were performed in a He3 cryostat at *0.3* K.

Using the top and bottom gates we can control the total charge density of the twisted bilayer and the interlayer electric field. We define the applied total electron density on the bilayer as $n_{tot} = (C_T V_T + C_B V_B)/e$, where $C_T$ and $C_B$ are the top and bottom gate capacitances per unit area, $V_T$ and $V_B$ are the top and bottom gate voltages, and $e$ is the electron charge. In a magnetic field, $B$, the relevant measure of charge density is the total filling factor $v_{tot} = n_{tot}(h/e)/B$, which is the number of filled Landau levels ($h/e$ is the magnetic flux quantum, where $h$ is Planck's constant). Applying antisymmetric gate voltages will impose an interlayer electric field that shifts charges between layers, causing them to have different filling factors. We present this experimental knob as the applied displacement field $D = (C_T V_T - C_B V_B)/2$ divided by the vacuum permittivity $\varepsilon_0$.

To establish the degree of interlayer coupling in our devices, we begin by measuring the quantum Hall effect. The quantum Hall effect is a sensitive probe of electron degeneracy and the underlying symmetries of the Landau levels; as such, the graphene quantum Hall effect is different for monolayers (*23, 24*), AB-stacked bilayers (*25*), and twisted bilayers (*26*). In a perpendicular magnetic field, each layer of the twisted bilayer will develop chiral 1d edge states which originate from the bulk Landau levels. The chirality, or direction of edge propagation, is determined by the filling factor sign in each layer and the magnetic field direction (Figure 1C). Changing the filling factor alters the number of edge states crossing the Fermi level, resulting in conductance steps quantized in units of $e^2/h$ times the state degeneracy.

Figure 1D shows a measurement of the 2-probe conductance, $G$, of a twisted bilayer device at $B = 1$ T. The device exhibits plateaus in conductance as $v_{tot}$ changes. The interlayer displacement field is kept at zero ($D = 0$) such that the filling factors of the top and bottom layers are equal ($v_{top} = v_{bottom} = v_{tot}/2$). The conductance jumps of $8\,e^2/h$ indicate 8-fold degenerate Landau levels due to spin, valley and layer symmetries. The observed quantum Hall sequence is double that of monolayer graphene (*26*). For example, the $G = 4\,e^2/h$ plateau occurs when each layer is in the monolayer graphene state $v_{top} = v_{bottom} = -2$, corresponding to a spin-degenerate edge state in each layer (Figure 1C). The sequence demonstrates that interlayer coupling is too weak to split the layer degeneracy. We conclude that the twist angle is large enough to model the system as two monolayer graphene sheets conducting in parallel (*26, 27*).

In low-disorder samples, electron exchange interactions can break the graphene spin-valley degeneracy, leading to quantum Hall ferromagnetism (*28-30*). We indeed observe such degeneracy breaking at higher field as a sequence of plateaus at all integer multiples of $e^2/h$ from $-4$ to $4$ ($B = 4$ T, Figure 1E). This can be explained by the exchange-driven breaking of spin-valley symmetry in each of the graphene layers, combined with the effects of displacement field. For example, at $v_{tot} = 0$, both layers are charge neutral and we observe an insulating state ($G = 0$), similar to the exchange-driven insulating state observed in neutral monolayer graphene (*14, 29, 30*).

Decreasing $v_{tot}$ from $0$ to $-1$, a small applied displacement field causes charge to be removed from the top layer preferentially. The result is a transition to a $1\,e^2/h$ plateau, which we explain as conduction through a hole-like edge state in the top layer while the bottom layer remains insulating (Figure 1E, left cartoon). Conversely, increasing $v_{tot}$ to $1$ preferentially adds charges to the bottom layer, resulting in an electron-like edge state with conductance of $1\,e^2/h$ (Figure 1E, right cartoon). We label these states by the filling factors on each layer as $(v_{bottom}, v_{top}) = (0,-1)$ and $(1,0)$. In monolayer graphene, the filling factor $v = \pm 1$ states are thought to be spin polarized due to quantum Hall ferromagnetism (*30, 31*).  At $v = 1$, the spin magnetic moment is aligned with the magnetic field; for the hole-like $v = -1$ edge state the spin is flipped since it originates from the bulk excited state.  If the same effect occurs in twisted bilayer graphene, it should be possible to create a pair of helical edge states with opposite chiralities and opposite spin polarizations by realizing coexisting $v = 1$ and $v = -1$ states.

We now explore the outcomes when the twisted bilayer is electron-hole doped such that the layers have edge states of opposite chiralities. Starting with each layer in the insulating state at charge neutrality *(0,0)*, we imbalance the bilayer with a displacement field such that the charge density from each layer is of equal magnitude but of opposite sign ($v_{top} = -v_{bottom}$, $v_{tot} = 0$, Figure 2A).  As the displacement field increases, the system first transitions to a conductive state of order $e^2/h$, and then transitions sharply to another insulating state at higher displacement fields (Figure 2B and 2C). Assuming that the transitions correspond to filling factor changes in each layer, we assign the conductive states to the *(±1, ∓1)* charge configurations, and the insulating states at higher *D* magnitudes to the *(±2, ∓2)* states.  We have consistently observed this conductance sequence in all large-twist bilayer graphene devices that display broken-symmetry states (9 devices in total), with *(±1, ∓1)* state conductances varying from *0.8* to *1.5 $e^2/h$*.

To verify the assignment of the *(±1, ∓1)* states, we study a wider range of edge state configurations away from $v_{tot} = 0$.  Figure 2D shows the 2-probe conductance as a function of $v_{tot}$ and displacement field.  The *(±1, ∓1)* states form clearly defined plateaus in the map (white dotted circle). We model the sequence by considering all possible combinations of filling factors in the graphene zeroth Landau level with broken spin-valley degeneracy.  The resulting map in Figure 2E matches the entire sequence of plateau transitions observed in the 2-probe conductance (Figure 2D) and 4-probe longitudinal resistance measurements (see Supplementary Fig. S1).  Furthermore, capacitance measurements on a different sample reveal that the bulk is insulating for each of the plateaus in the map, as expected for quantum Hall states (Supplementary Fig. S2).  The consistency of the model with the observed plateaus supports the assignment of the conductive $v_{tot} = 0$ states to the *(±1, ∓1)* filling factor configurations.

The conductances of nearly all the filling factor configurations are determined by the total filling factor: $G = v_{tot} e^2/h$.  Noticeably, only the *(±1, ∓1)* states depart from this pattern.  When $v_{bottom}$ and $v_{top}$ have the same sign, this formula follows directly from the parallel conductance contributions of quantum Hall edge states in each layer.  But for electron-hole bilayer combinations, such as the *(+2,-2)* or *(+2,-1)* states, this equation implies that conductance contributions from each layer can cancel.  For this to occur, there must be a backscattering process that couples the edge states between layers (bottom cartoon of Figure 2A) (*26*).  Moreover, temperature dependence of the insulating *(±2, ∓2)* states suggests that this backscattering leads to a complete transport gap (Supplementary Fig. S5).  Interlayer backscattering requires tunneling between the closely spaced layers, which may be enhanced at the edge even if it is suppressed in the bulk.  In contrast, the same backscattering process is nearly absent in the *(±1, ∓1)* states, resulting in a conductive plateau of order $e^2/h$ for a device with greater than *5* μm long edges.

We now show that the *(±1, ∓1)* states conduct through counter-propagating edge modes by measuring the nonlocal voltage response in the same device. Figure 3B depicts the measurement schematic, where the voltage $V_{NL}$ is measured between adjacent contacts far away from the source and drain electrodes. The nonlocal resistance $R_{NL}$ is defined as $V_{NL}/I_M$, where $I_M$ is the measured source-drain current, and is presented as function of $v_{tot}$ and displacement field in Figure 3C. The nonlocal resistances of the *(±1, ∓1)* states are *10* to *1000* times larger than the other conductive states (white dots outline the *(±1, ∓1)* states). Figure 3D shows a comparison of the nonlocal and 2-probe resistance ($R_{2probe}$) as $v_{tot}$ is tuned through the *(+1,-1)* state. When $R_{2probe}$ exhibits a quantum Hall plateau, the value of $R_{NL}$ is flat and close to zero (*1-100* Ω), since the voltage drop along a chiral edge state is zero. During plateau transitions, a small nonlocal peak can be observed as the bulk becomes conductive. The bulk contribution to the signal is small because the nonlocal voltage drop in a diffusive conductor will fall off exponentially away from the source-drain electrodes as $V_{xx} \sim \rho_{xx} \exp(-L/W)$, where $\rho_{xx}$ is the local resistivity, and *L* and *W* are the device length and width. In contrast to the weak bulk response, the strong nonlocal resistance of the *(±1, ∓1)* states signifies that current flows predominately via counterpropagating modes along the edges of the device.

Based on the transport data collected—the mapping of the quantum Hall plateau sequence and the edge state nonlocal signal—we conclude that at filling factors *(±1, ∓1)* conduction occurs through 1d edge modes corresponding to quantum Hall states with opposite chiralities (middle cartoon, Figure 2A). Backscattering between the two counter-propagating modes is strongly suppressed, resulting in a highly conductive 1d transport channel with conductance ranging from *0.8* to *1.5* $e^2/h$ for devices with edge lengths varying from *0.2* to *16* μm (details in Supplementary Fig. S9). This is contrasted with the spin-degenerate *(±2, ∓2)* states, where interlayer scattering leads to insulating behavior in the same devices (Figure 2A). A simple explanation for the difference is that the counter-propagating modes of the *(±1, ∓1)* states have opposite spin polarizations, which are the expected exchange-driven ground states for monolayer graphene at $v = \pm 1$(30, 31). When the spin-wavefunctions on each layer are orthogonal, interlayer tunneling processes are forbidden and the edge states are protected from backscattering. The result is a pair of helical edge states in the *(±1, ∓1)* electron-hole bilayer.

The expected conductance of the helical edge states is *2* $e^2/h$ (resulting from the combined transport through the two counterpropagating edges) when backscattering is completely suppressed. Even though our measured conductance is high given the device lengths (e.g. near *1* $e^2/h$ for > *5* μm edges), all measured devices have a conductance below *2* $e^2/h$ in the *(±1, ∓1)* states. A significant reduction in the 2-probe measurement is due to contact resistance from the electrode-edge state interface (see Supplementary Materials). To avoid the effects of contact resistance, we measure the 4-probe resistance of the *(+1,-1)* states as a function of magnetic field in both local ($R_{xx}$) and nonlocal configurations ($R_{NL}$) (Figure 3E). Above *1.5* T, $R_{NL}$ increases slowly until it saturates at high fields, while $R_{xx}$ decreases to approach a similar value, despite the two measurements probing edges of very different lengths. Moreover, the measurements approach $h/4e^2$—the expected value for ballistic counter-propagating edge states that fully equilibrate at the contacts. The convergence of $R_{xx}$ and $R_{NL}$ suggests that backscattering in the helical *(±1, ∓1)* states decreases steadily with increasing magnetic field, causing a length-independent edge segment resistance of $h/e^2$. This is consistent with our 2-probe measurements: the short and long edges of the device have very similar conductance at *8* T (*1.3* vs *1.1* $e^2/h$ respectively) despite a *3x* difference in length and similar contact resistances (see Supplementary Fig. S9).

We now turn to the low field regime of the *(±1, ∓1)* states (Figure 3E). At zero magnetic field, the nonlocal resistance is insignificant (*1-10* Ω); as the magnetic field rises to *1.5* T, $R_{NL}$ sharply increases by a factor of *100*. This coincides with the emergence of clearly distinguished plateaus at *(±1, ∓1)* in both the $R_{NL}$ and $R_{2probe}$ maps (Supplementary Fig. S10). We interpret the sharp increase in $R_{NL}$ as the onset of conduction in the helical edge states at *1.5* T, a comparatively low field that is encouraging for future efforts to engineer topological superconductivity in this helical conductor (*1, 2*).

One unique advantage of building a helical 1-dimensional conductor from quantum Hall edge states is the possibility of extending the system to fractional edge states, since monolayer graphene exhibits the fractional quantum Hall effect (*32, 33*). As a promising step in this direction, we have observed the fractional quantum Hall effect in twisted bilayer graphene devices by measuring at higher magnetic fields. Figure 4A shows a 2-probe conductance measurement of a different twisted bilayer device at *31* T, where clear plateaus are observed at multiples of *1/3* $e^2/h$. From the location of the line cuts in the hole-hole bilayer regime (Figure 4B), we infer that the fractional plateaus are due to the parallel conduction through an integer edge state in one layer and a fractional edge state in the other. Interestingly, we also observe fractional plateaus in the electron-hole bilayer regime. Figure 4B shows a line cut from *(-2,0)* to *(-2,+2)* for another device, where clear fractional plateaus are observed which we identify with the *(+2/3,-2)* and *(+5/3,-2)* states. The conductance is also given by the sum of the filling factors, which suggests again that inter-layer backscattering is present in these electron-hole bilayers. In this case, the fractional edge state in one layer is able to reduce the conductance in the other layer by a fractional multiple of $e^2/h$. Despite clear evidence for fractional edge states in this electron-hole bilayer regime, contact resistances are too high at these magnetic fields to probe possible helical states near filling factors *(±1, ∓1)*. Our preliminary results with new fabrication techniques indicate that similar fractional quantum Hall states can be realized in twisted bilayer graphene at much lower magnetic fields ( < *9* T), which bodes well for future attempts to achieve a fractional quantum spin Hall state in this system (*6-8*).

## Acknowledgements

We acknowledge helpful discussions with L. Levitov, L. Fu, and V. Fatemi. We also acknowledge fabrication help from D. Wei, G. H. Lee, S. H. Choi and Y. Cao. This work has been primarily supported by the National Science Foundation (DMR-1405221) for device fabrication, transport and data analysis (J.D.S.Y., J.Y.L., P.J.H.), with additional support from the NSS Program, Singapore (J.Y.L.). This research has been funded in part by the Gordon and Betty Moore Foundation's EPiQS Initiative through Grant GBMF4541 to P.J.H. The capacitance measurements have been supported in part by the Gordon and Betty Moore Foundation Grant GBMF2931 to R.C.A. and by the STC Center for Integrated Quantum Materials, NSF Grant (DMR-1231319) (A.F.Y., B.H. and R.C.A.). This work made use of the Materials Research Science and Engineering Center Shared Experimental Facilities supported by the National Science Foundation (DMR-0819762) and of Harvard's Center for Nanoscale Systems, supported by the NSF (ECS-0335765). Some measurements were performed at the National High Magnetic Field Laboratory, which is supported by NSF Cooperative Agreement DMR-1157490 and the State of Florida.

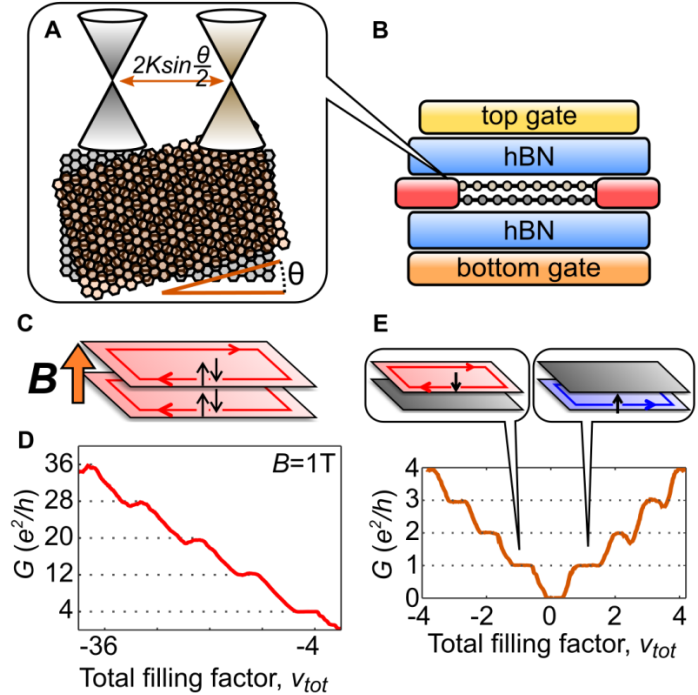

**Fig. 1.** Quantum Hall effect in twisted bilayer graphene with broken symmetry states. (A) Stacking two graphene layers with a relative twist decouples the Dirac cones from each layer via a large momentum mismatch. (B) Device schematic of twisted bilayer graphene encapsulated in hBN with dual-gates. Contact electrodes depicted in red. (C) Cartoon of twisted bilayer quantum Hall edge states when both layers are at filling factor -2. (D) 2-probe conductance of a twisted bilayer graphene device at $B = 1$ T as a function of $\nu_{tot}$. A contact resistance has been subtracted to fit the $\nu_{tot} = -4$ plateau to $4\,e^2/h$. (E) 2-probe conductance of the same device at $B = 4$ T showing broken-symmetry states. Contact resistances have been subtracted from the negative and positive $\nu_{tot}$ sides of the data. Cartoons depict proposed edge state configurations in the $(0,-1)$ and $(+1,0)$ states. Note that this trace is taken at a small D in order to properly observe all the integer steps (see color map in Fig. 2E).

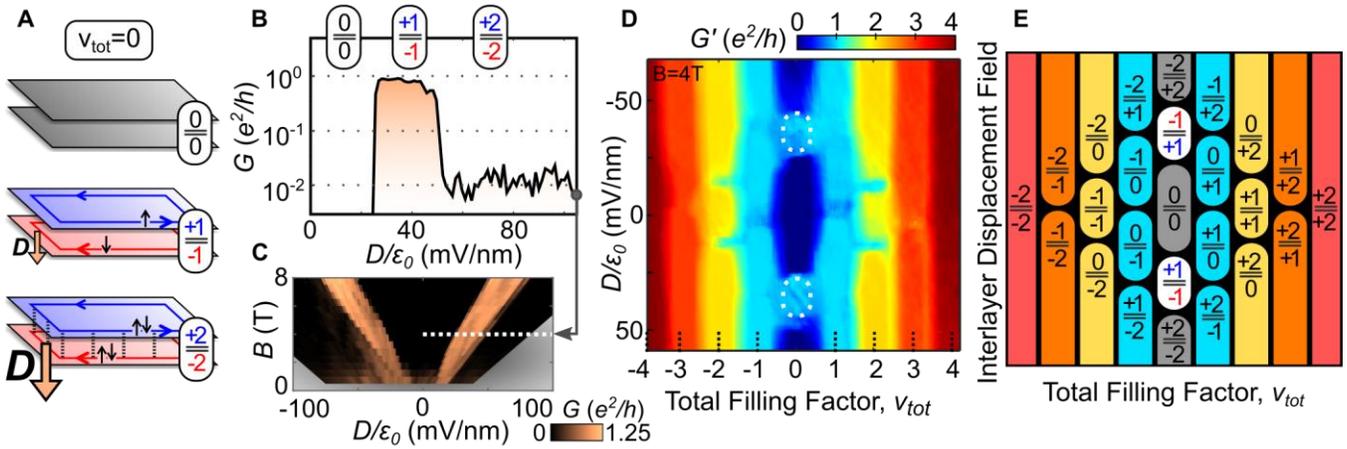

**Fig. 2.** Transport in graphene electron-hole bilayers. (A) Cartoons depicting edge state configurations with $v_{top} = -v_{bottom}$. (B) Conductance for $v_{tot} = 0$ as a function of displacement field at $B = 4$ T. (C) Magnetic field dependence of $v_{tot} = 0$ line. (D) 2-probe conductance map, $G'$, as function of $v_{tot}$ and $D$. Conductance is given by $\frac{e^2}{h}v_{tot}$ for all configurations except for the $(\pm 1, \mp 1)$ states. Contact resistances have been subtracted from the positive and negatives sides of the data to fit the $v_{tot} = \pm 1$ plateaus. (E) Schematic map of possible filling factor combinations.

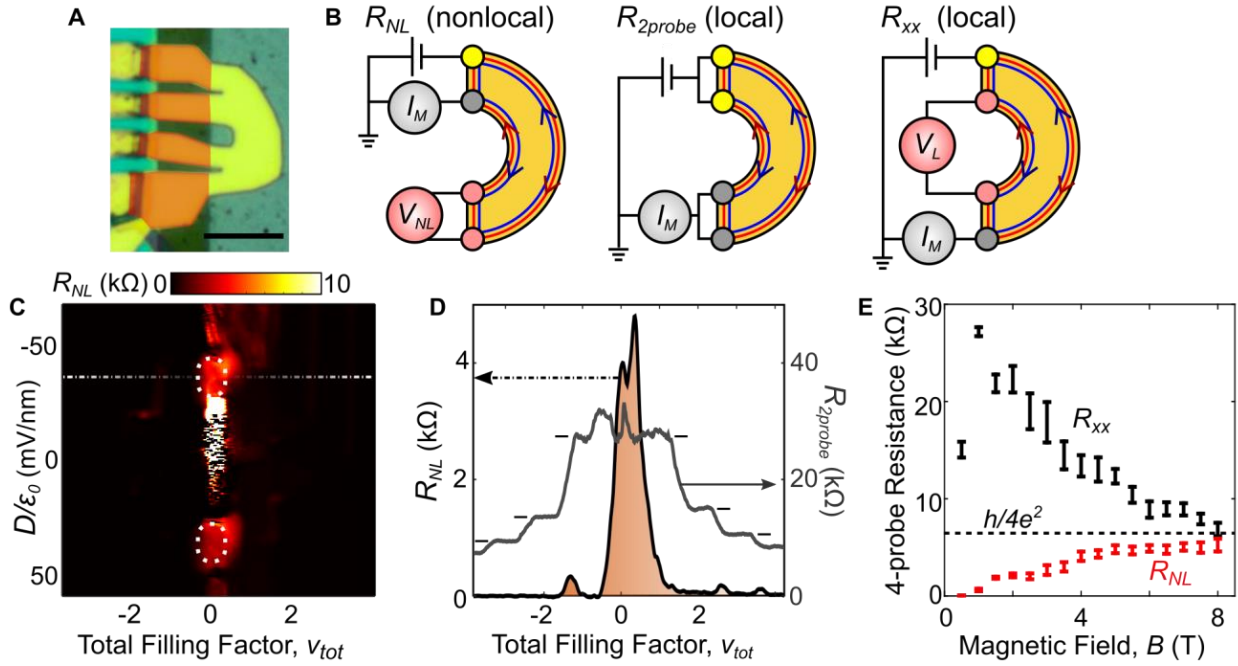

**Fig. 3.** Nonlocal measurements of helical edge states. (A) Optical image of 4-probe device with 5 μm scale bar. Graphite leads are highlighted in red. (B) Schematic of different measurement configurations for the 4-probe device. (C) Nonlocal resistance as a function of $v_{tot}$ and displacement field, $D$. Dashed white circles highlight $(\pm 1, \mp 1)$ states. Axis ranges are identical to Figure 2D. In the $(0,0)$ insulating state, $R_{NL}$ fluctuates strongly due to low current signals near the noise limit (bright white features). (D) Nonlocal resistance (black line, left axis) compared to 2-probe resistance (grey line, right axis) of constant $D$ line cut through $(+1,-1)$ state (dashed line in Figure 3C). (E) Magnetic field dependence of $R_{NL}$ and $R_{xx}$ in $(+1,-1)$ state. In the ballistic limit, each edge segment has resistance $h/e^2$, leading to a 4-probe resistance of $h/4e^2$.

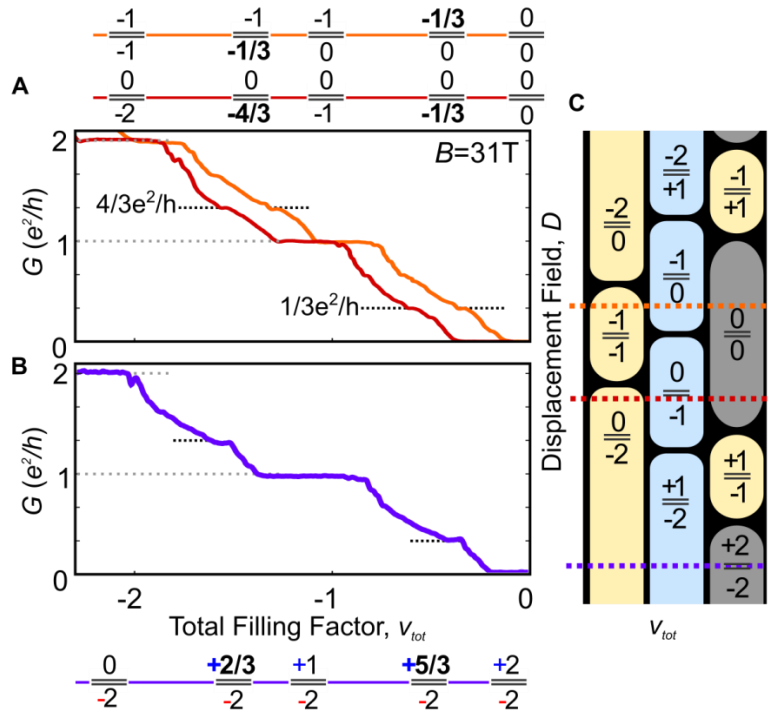

**Fig. 4.** Fractional quantum Hall effect in twisted bilayer graphene at $B = 31T$. (A) 2-probe conductance measurement as a function of $v_{tot}$ for two different displacement field values (measurements offset horizontally for clarity). A contact resistance has been subtracted to fit the $v_{tot} = -1$ plateau. Proposed filling factor sequence of observed plateaus is shown on top. (B) 2-probe conductance of a different device in the electron-hole bilayer case showing electron-hole bilayer fractions. Proposed filling factor sequence is shown at bottom. (C) Filling factor map showing location of line cuts.

# References

1. M. Z. Hasan, C. L. Kane, Topological insulators. *Rev. Mod. Phys.* **82**, 3045 (2010).
2. X.-L. Qi, S.-C. Zhang, Topological insulators and superconductors. *Rev. Mod. Phys.* **83**, 1057 (2011).
3. M. Konig, S. Wiedmann, C. Brune, A. Roth, H. Buhmann, L. W. Molenkamp, X. L. Qi, S. C. Zhang, Quantum spin hall insulator state in HgTe quantum wells. *Science* **318**, 766 (2007).
4. I. Knez, R.-R. Du, G. Sullivan, Evidence for Helical Edge Modes in Inverted InAs/GaSb Quantum Wells. *Physical Review Letters* **107**, 136603 (2011).
5. V. Mourik, K. Zuo, S. M. Frolov, S. R. Plissard, E. P. A. M. Bakkers, L. P. Kouwenhoven, Signatures of Majorana Fermions in Hybrid Superconductor-Semiconductor Nanowire Devices. *Science* **336**, 1003 (2012).
6. N. H. Lindner, E. Berg, G. Refael, A. Stern, Fractionalizing Majorana Fermions: Non-Abelian Statistics on the Edges of Abelian Quantum Hall States. *Physical Review X* **2**, 041002 (2012).
7. M. Cheng, Superconducting proximity effect on the edge of fractional topological insulators. *Physical Review B* **86**, 195126 (2012).
8. D. J. Clarke, J. Alicea, K. Shtengel, Exotic non-Abelian anyons from conventional fractional quantum Hall states. *Nat Commun* **4**, 1348 (2013).
9. G. M. Gusev, E. B. Olshanetsky, Z. D. Kvon, A. D. Levin, N. N. Mikhailov, S. A. Dvoretsky, Nonlocal Transport Near Charge Neutrality Point in a Two-Dimensional Electron-Hole System. *Physical Review Letters* **108**, 226804 (2012).
10. F. Nichele, A. N. Pal, P. Pietsch, T. Ihn, K. Ensslin, C. Charpentier, W. Wegscheider, Insulating State and Giant Nonlocal Response in an In As/Ga Sb Quantum Well in the Quantum Hall Regime. *Physical Review Letters* **112**, 036802 (2014).
11. D. A. Abanin, P. A. Lee, L. S. Levitov, Spin-filtered edge states and quantum Hall effect in graphene. *Phys Rev Lett* **96**, 176803 (2006).
12. H. A. Fertig, L. Brey, Luttinger Liquid at the Edge of Undoped Graphene in a Strong Magnetic Field. *Phys. Rev. Lett.* **97**, 116805 (2006).
13. P. Maher, C. R. Dean, A. F. Young, T. Taniguchi, K. Watanabe, K. L. Shepard, J. Hone, P. Kim, Evidence for a spin phase transition at charge neutrality in bilayer graphene. *Nature Physics* **9**, 154 (2013).
14. A. F. Young, J. D. Sanchez-Yamagishi, B. Hunt, S. H. Choi, K. Watanabe, T. Taniguchi, R. C. Ashoori, P. Jarillo-Herrero, Tunable symmetry breaking and helical edge transport in a graphene quantum spin Hall state. *Nature* **505**, 528 (2014).
15. L. Wang, I. Meric, P. Huang, Q. Gao, Y. Gao, H. Tran, T. Taniguchi, K. Watanabe, L. Campos, D. Muller, One-dimensional electrical contact to a two-dimensional material. *Science* **342**, 614 (2013).
16. P. J. Zomer, M. H. D. Guimarães, J. C. Brant, N. Tombros, B. J. van Wees, Fast pick up technique for high quality heterostructures of bilayer graphene and hexagonal boron nitride. *Applied Physics Letters* **105**, 013101 (2014).
17. J. M. B. Lopes dos Santos, N. M. R. Peres, A. H. Castro Neto, Graphene Bilayer with a Twist: Electronic Structure. *Physical Review Letters* **99**, 256802 (2007).
18. A. Luican, G. Li, A. Reina, J. Kong, R. Nair, K. Novoselov, A. Geim, E. Andrei, Single-Layer Behavior and Its Breakdown in Twisted Graphene Layers. *Physical Review Letters* **106**, 126802 (2011).
19. R. de Gail, M. O. Goerbig, F. Guinea, G. Montambaux, A. H. Castro Neto, Topologically protected zero modes in twisted bilayer graphene. *Physical Review B* **84**, 045436 (2011).
20. M.-Y. Choi, Y.-H. Hyun, Y. Kim, Angle dependence of the Landau level spectrum in twisted bilayer graphene. *Physical Review B* **84**, 195437 (2011).
21. P. Moon, M. Koshino, Energy spectrum and quantum Hall effect in twisted bilayer graphene. *Physical Review B* **85**, 195458 (2012).
22. A. S. Mayorov, R. V. Gorbachev, S. V. Morozov, L. Britnell, R. Jalil, L. A. Ponomarenko, P. Blake, K. S. Novoselov, K. Watanabe, T. Taniguchi, A. K. Geim, Micrometer-Scale Ballistic Transport in Encapsulated Graphene at Room Temperature. *Nano Letters* **11**, 2396 (2011).
23. Y. Zhang, Y. W. Tan, H. L. Stormer, P. Kim, Experimental observation of the quantum Hall effect and Berry's phase in graphene. *Nature* **438**, 201 (2005).

24. K. S. Novoselov, A. K. Geim, S. V. Morozov, D. Jiang, M. I. Katsnelson, I. V. Grigorieva, S. V. Dubonos, A. A. Firsov, Two-dimensional gas of massless Dirac fermions in graphene. *Nature* **438**, 197 (2005).
25. K. S. Novoselov, E. McCann, S. V. Morozov, V. I. Fal'ko, M. I. Katsnelson, U. Zeitler, D. Jiang, F. Schedin, A. K. Geim, Unconventional quantum Hall effect and Berry's phase of 2π in bilayer graphene. *Nature Physics* **2**, 177 (2006).
26. J. D. Sanchez-Yamagishi, T. Taychatanapat, K. Watanabe, T. Taniguchi, A. Yacoby, P. Jarillo-Herrero, Quantum Hall Effect, Screening, and Layer-Polarized Insulating States in Twisted Bilayer Graphene. *Physical Review Letters* **108**, 076601 (2012).
27. H. Schmidt, T. Lüdtke, P. Barthold, R. J. Haug, Mobilities and scattering times in decoupled graphene monolayers. *Physical Review B* **81**, 121403(R) (2010).
28. Y. Zhang, Z. Jiang, J. P. Small, M. S. Purewal, Y. W. Tan, M. Fazlollahi, J. D. Chudow, J. A. Jaszczak, H. L. Stormer, P. Kim, Landau-level splitting in graphene in high magnetic fields. *Physical Review Letters* **96**, 136806 (2006).
29. J. G. Checkelsky, L. Li, N. P. Ong, Zero-Energy State in Graphene in a High Magnetic Field. *Phys. Rev. Lett.* **100**, 206801 (2008).
30. A. F. Young, C. R. Dean, L. Wang, H. Ren, P. Cadden-Zimansky, K. Watanabe, T. Taniguchi, J. Hone, K. L. Shepard, P. Kim, Spin and valley quantum Hall ferromagnetism in graphene. *Nature Physics* **8**, 550 (2012).
31. F. Amet, J. R. Williams, K. Watanabe, T. Taniguchi, D. Goldhaber-Gordon, Selective Equilibration of Spin-Polarized Quantum Hall Edge States in Graphene. *Physical Review Letters* **112**, 196601 (2014).
32. X. Du, I. Skachko, F. Duerr, A. Luican, E. Y. Andrei, Fractional quantum Hall effect and insulating phase of Dirac electrons in graphene. *Nature* **462**, 192 (2009).
33. K. I. Bolotin, F. Ghahari, M. D. Shulman, H. L. Stormer, P. Kim, Observation of the fractional quantum Hall effect in graphene. *Nature* **462**, 196 (2009).
34. E. McCann, Asymmetry gap in the electronic band structure of bilayer graphene. *Physical Review B* **74**, (2006).
35. T. Taychatanapat, P. Jarillo-Herrero, Electronic Transport in Dual-Gated Bilayer Graphene at Large Displacement Fields. *Physical Review Letters* **105**, (2010).

# Supplementary Information for

## Observation of Helical Edge States and Fractional Quantum Hall Effect in a Graphene Electron-hole Bilayer

J. D. Sanchez-Yamagishi, J. Y. Luo, A. F. Young, B. Hunt, K. Watanabe, T. Taniguchi, R. C. Ashoori, P. Jarillo-Herrero

Correspondence to: pjarillo@mit.edu

Fabrication Methods

All twisted bilayer graphene devices were made using a dry-transfer process (*15, 16*) to create a van der Waals heterostructure consisting of hBN-graphene-graphene-hBN layers. The devices all are dual-gated with a top and bottom gate electrode. For the device leads which contact the dual-gated region, we used two different approaches: graphite contacts and gate-tuned contacts. Graphite contacts provide the simplest approach, where graphite is used as the electrode material to contact the twisted bilayer graphene. The advantage of graphite is that it has a work function similar to graphene, and hence does not cause strong local doping at the contact interface. As a result, graphite can provide good contact to both electron-doped and hole-doped graphene layers, even at high magnetic fields. Graphite contacts are used for the device data presented in Figure 1, 2 and 3 of the main text. An alternative method is to use local gate electrodes which separately gate the twisted bilayer graphene outside the primary dual gated region. The advantage of this method is that contact resistances can be controllably reduced to sub-100Ω range, even at high magnetic fields, but can only be used to contact well either electron-doped or hole-doped graphene (Further discussion in section: Contact resistances and layer-selective contacts).

A summary of the typical fabrication steps are as follows:

1. Hexagonal boron nitride and graphene flakes are exfoliated onto Piranha + HF cleaned Si/SiO2 chips. Flakes are identified by optical microscopy and checked for cleanliness in an atomic force microscope (AFM).
2. Flakes are picked up and transferred using a transparent polymer stamp made from either polypropylene carbonate (PPC)(*15*) or polycarbonate (PC)(*16*). The top hBN crystal is picked up first using the stamp, and then subsequently the hBN is used to pick up two graphene flakes and then the bottom hBN. The graphene flakes are rotated so that natural edges are mis-aligned to avoid producing a low twist angle sample. For graphite contacts, a layer of thin graphite (<20nm thick) is also picked up that overlaps with the graphene layers to provide electrical contact.
3. The complete stack consisting from top to bottom of hBN-graphene-graphene-hBN is then transferred onto a bottom gate electrode made of either graphite or a thin layer of PdAu 40:60 alloy (~20nm). The device discussed in the main text is made on a graphite bottom gate. The stack is then measured in an AFM to check for regions free of bubbles and ripples in the stack.  To increase flatness, the stack is also heat cleaned in forming gas (Ar:H2).  For the specific device discussed in Figure 1, 2 & 3 of the main text, the device was heat cleaned at 550C for 30mins to redistribute trapped residue between the flakes.
4. An isolated top gate electrode is made using electron beam lithography and evaporating Cr:Au (1nm:30-50nm).  The device geometry is then defined using reactive ion etching in a gas mixture of CHF3:O2:Ar, where the metal topgate and additional pmma is used as an etch mask.
5. A bridge contact is made to the metal topgate by depositing crosslinked PMMA and then Cr:Au.  The crosslinked PMMA avoids shorting to the exposed graphene at the edges of the device.
6. Final edge contacts are made to the device by an additional reactive ion etch step and then subsequently evaporating Cr:Au contacts using the same PMMA mask.  A rotation stage set at a 15 degree angle is used during the evaporation to make sure the metal properly coats the sidewalls of the device to ensure good 1d edge contact.

Contact resistance subtraction

A 2-probe measurement of quantum Hall edge states will include an extra resistance from the electrodes leading up to the measurement area of the device.  We correct for this contact resistance by extracting the resistance offset of a conductance plateau from its expected quantized value, for example, by fitting the $v=-1$ plateau to $1\ e^2/h$.  We then take this contact resistance and subtract it from the entire measurement.  The procedure is considered valid if a single contact resistance subtraction causes all other conductance plateaus to match an integer multiple of $e^2/h$ (see for example the data in Figure 1D of the main text).  We find that the contact resistance does change for positive and negative values of $v_{tot}$, due to the formation of pn junctions at the electrode-graphene interface (more discussion below).  As such, we perform separate contact resistance corrections for the negative and positive sides of a $v_{tot}$ sweep (this is the case for the data in Figure 1E and Figure 2D of the main text).  At *4*T, the device discussed in Figures 1-3 of the main text has a contact resistance of *0.6* kΩ for negative $v_{tot}$ and *2.0* kΩ for positive $v_{tot}$.  Since the contact resistance effects for the helical ($\pm 1, \mp 1$) are currently unknown, the measurement for the electron-hole bilayer configurations in Figure 2B & 2C of the main text are presented in the raw uncorrected form.

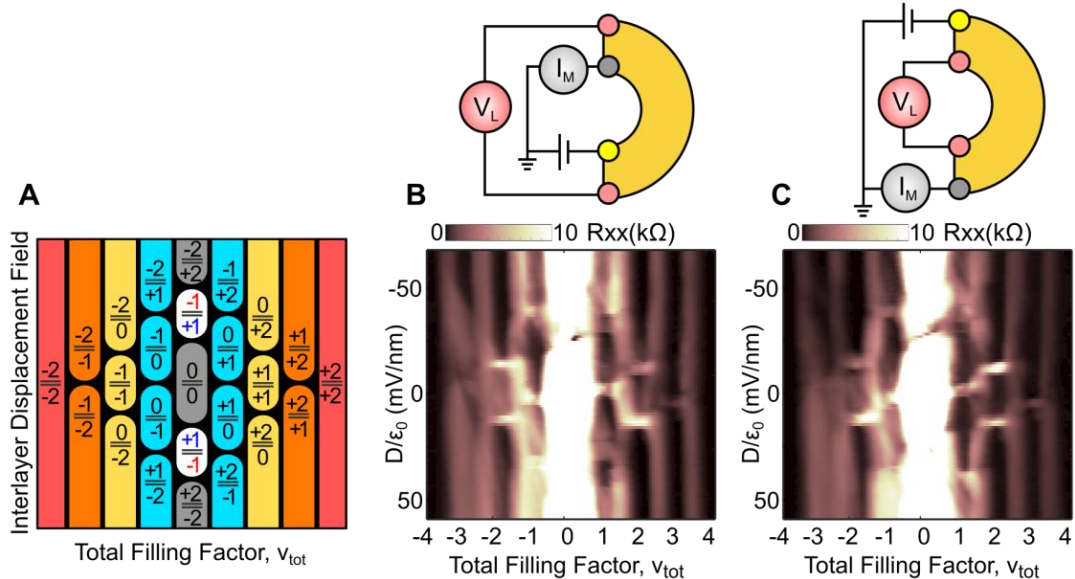

**Fig. S1.**
4-probe resistance maps show filling factor transitions. Data is for sample W presented in Figures 1-3 of the main text. Measurement is performed in two configurations (B) and (C), with the transitions matching the expected state sequence described in Figure 2E of the main text (reproduced here) and matches the plateau transitions observed in the 2-probe conductance data (Figure 2D main text). Measurement is at $B = 4$T.

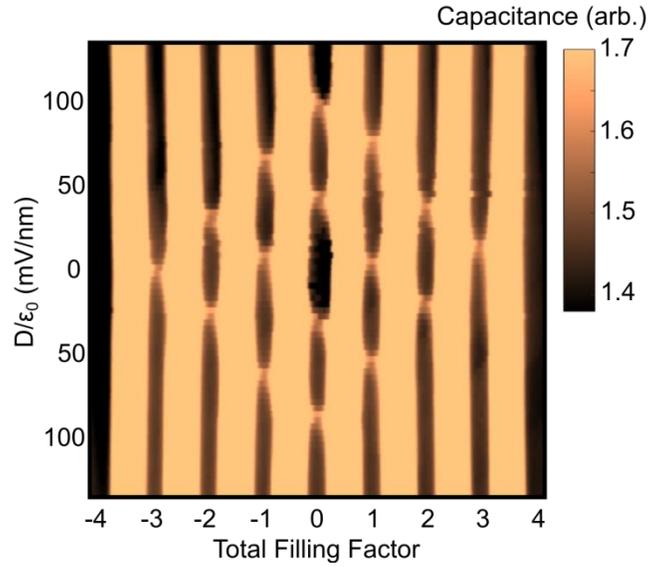

**Fig. S2**

Capacitance measurement on a twisted bilayer graphene device showing bulk state transitions (Sample O). Measurement signal is proportional to the device capacitance from the graphene bilayer to both gate electrodes. Low signal (black) corresponds to gapped/insulating states. High signal (orange) corresponds to high density of states/conductive states. Sequence of transitions matches 2-probe data and model presented in Figures 2D & 2E of the main text. Measurement is at $B = 18$ T.

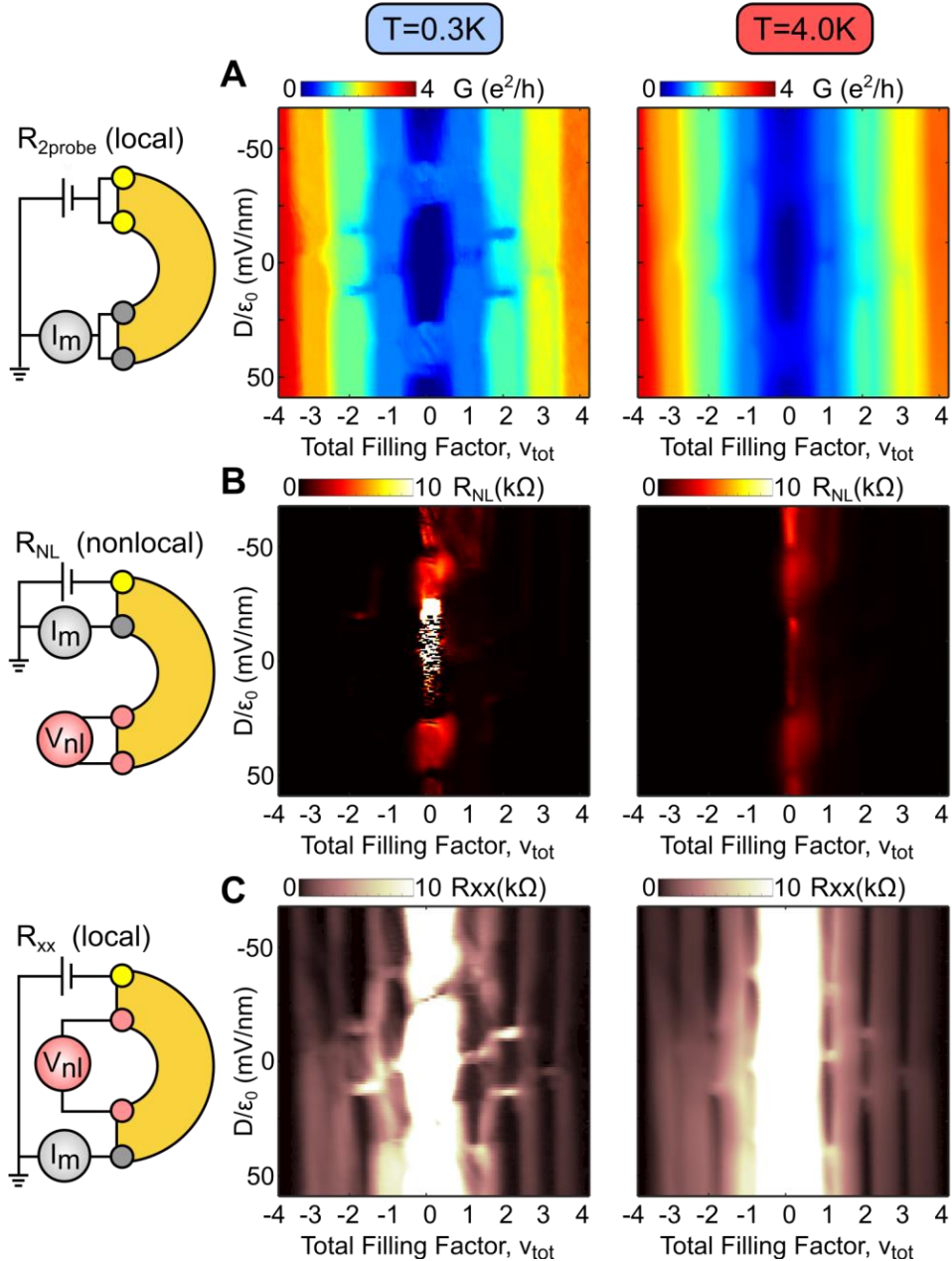

**Fig. S3**

Comparison of transport measurements at base (*0.3* K) and elevated temperatures (*4* K). Data is for sample W presented in Figures 1-3 of the main text. Columns from left to right correspond to *0.3* K and *4.0* K temperatures, respectively. Rows from top to bottom are the 2-probe conductance, nonlocal resistance, and local 4-probe resistance. Temperature causes a smooth broadening of all transport features.

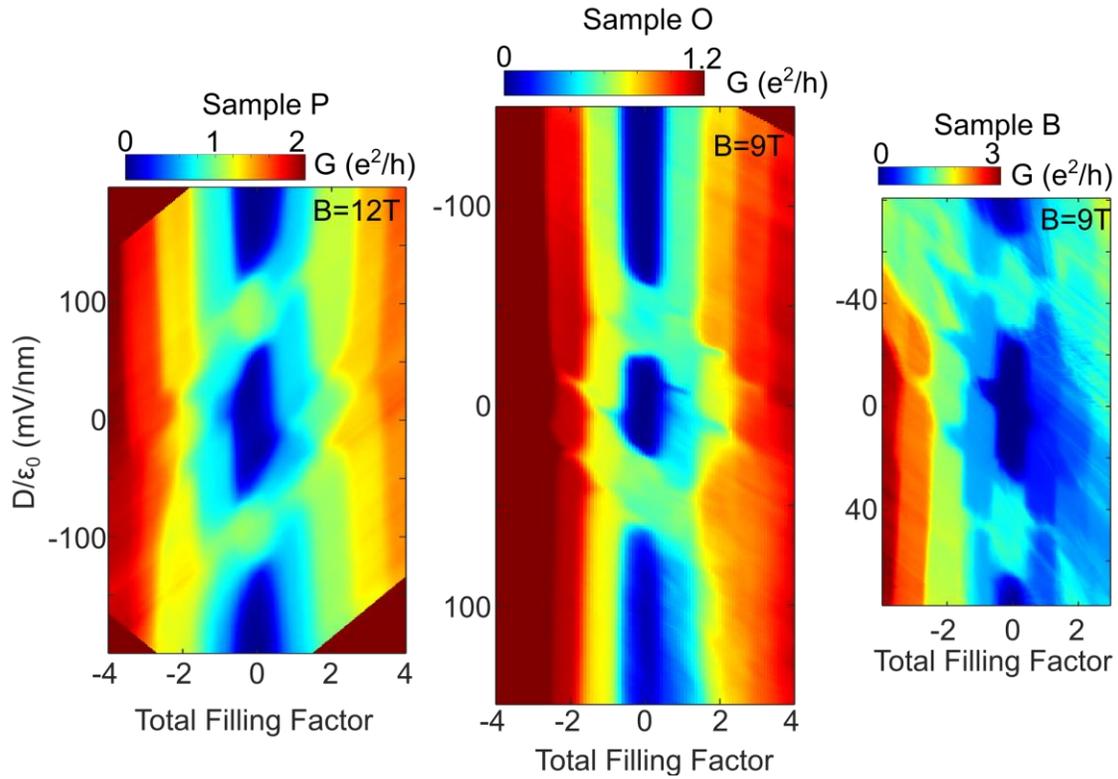

**Fig. S4**

Conductance maps for different devices as a function of displacement field and total filling factor. All show a similar pattern of plateau transitions as presented in Figure 2D of the main text. Departures from the expected plateau structure are understood as originating from gate-dependent contact resistance effects in these particular measurement geometries. Data is the raw 2-probe conductance without contact resistance corrections.

Temperature dependence of the electron-hole (+2,-2) state

The (±2, ∓2) states are characterized by deep insulating behavior that increases with magnetic field. Figure S5A shows the temperature dependence of the resistance in the (+2,-2) state at different electric and magnetic fields. The resistance shows an activated temperature dependence: $R \sim R_0 \exp(\Delta/2k_B T)$. The extracted activated gap increases with increased magnetic field (Figure S5B). The observation of thermally activated behavior suggests that a hybridization gap is developing between the counter-propagating (±2, ∓2) edge states originating from interlayer tunneling at the edge.

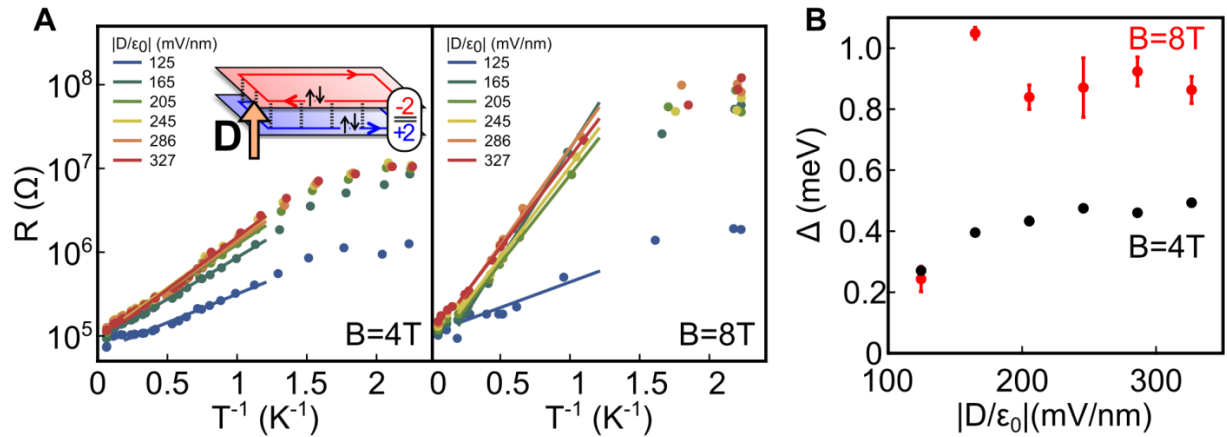

**Fig. S5**

Temperature dependence of the (+2,-2) insulating state at *4* T and *8* T shows an activated dependence, suggesting a full hybridization gap at the sample edge between the counter-propagating edge states. Data is for Sample W presented in Figures 1-3 of the main text. (A) 2-probe resistance as a function of inverse temperature for (+2,-2) state at *4* T and *8* T. Lines show fit to activated temperature dependence $R \sim R_0 \exp(\Delta/2k_B T)$. (B) Extracted activated gaps as a function of displacement field.

Transport properties of the (±1,∓1) states: temperature, bias, magnetic field and other devices

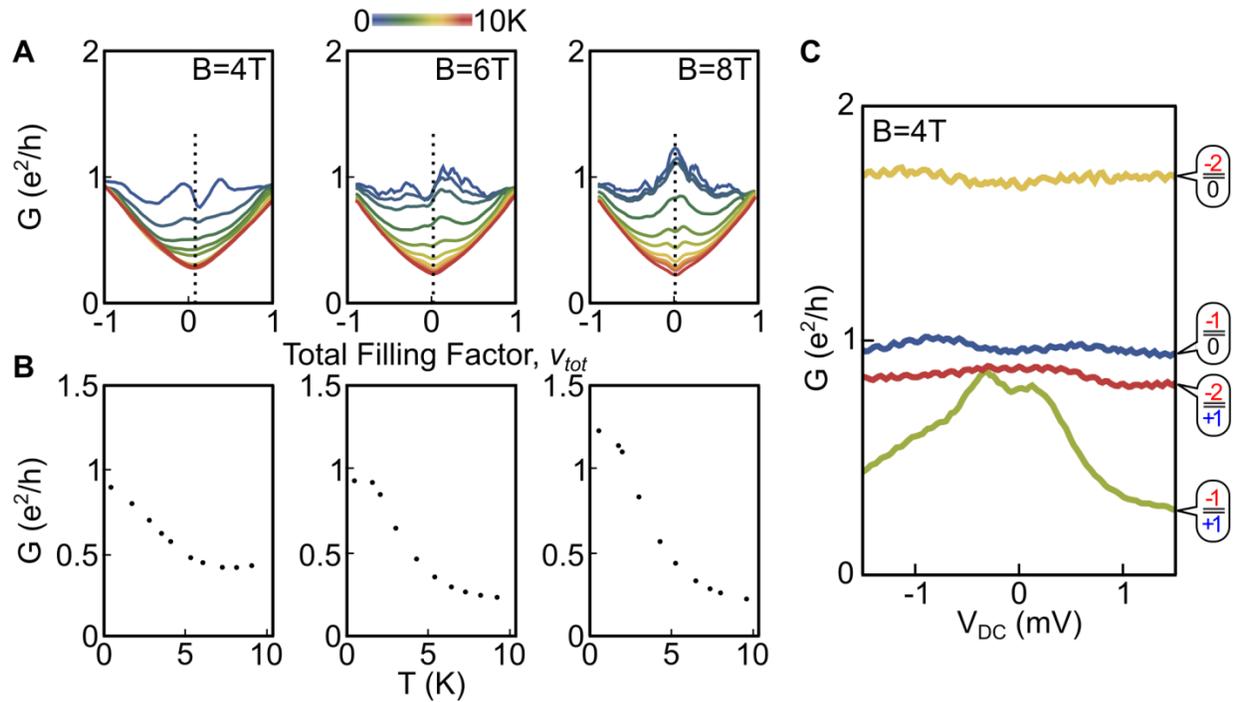

**Fig. S6**

Temperature and bias dependence of helical states. Data is for sample W presented in Figures 1-3 of the main text. (A) Conductance as a function of total filling factor at different temperatures (*0.3* K to *10* K) and magnetic fields. (*+1,-1*) state is centered at total filling factor zero. (B) Conductance of the (*+1,-1*) state as a function of temperature. Data is taken from points intersecting the vertical dotted line in the top datasets. The conductance through the helical edge states increases with higher magnetic fields and lower temperatures. (C) DC voltage bias dependence of differential conductance for different filling factor configurations. All measurements show a flat bias dependence except the helical (*+1,-1*) state, which shows an overall decrease in differential conductance with increasing voltage bias.

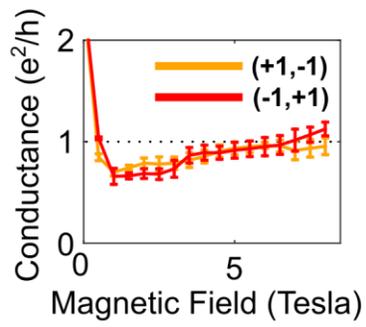

**Fig. S7**

Magnetic field dependence of 2-probe conductance in ($\pm1, \mp1$) states as a function of magnetic field. Data is for sample W presented in Figures 1-3 of the main text. Conductance steadily increases with magnetic field. Data is from Sample W discussed in Figures 1, 2 and 3 in the main text.

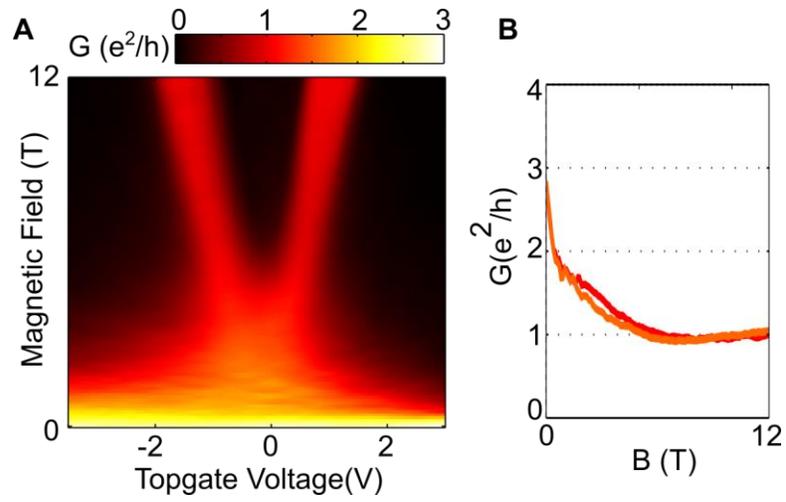

**Fig. S8**

Magnetic field dependence of $v_{tot}= 0$ line (A) and ($\pm 1$, $\mp 1$) states (B) for Sample P. Device shows a similar behavior to that presented in Figure 2C of the main text.

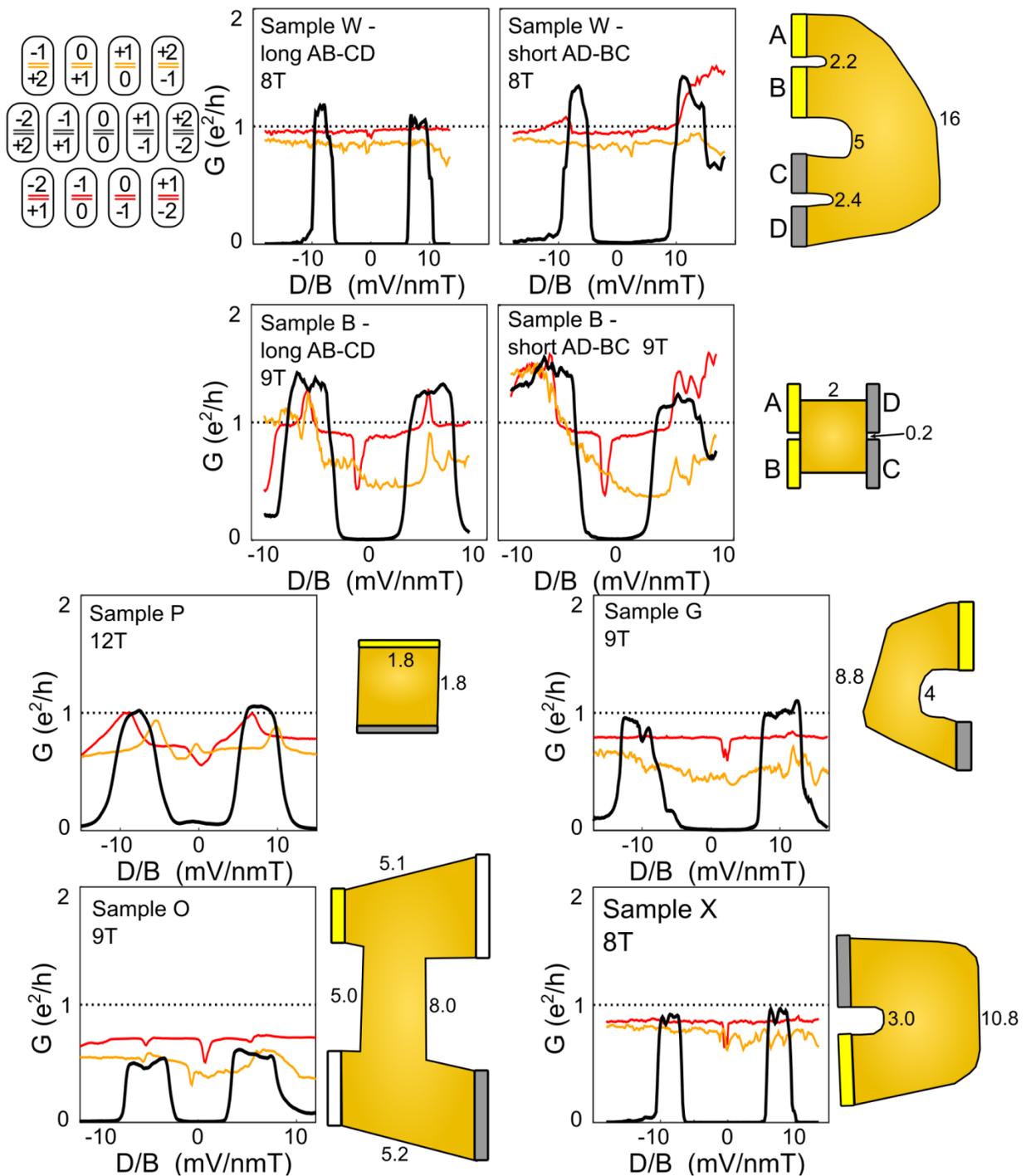

**Fig. S9**

Raw conductance of $v_{tot}= 0$ line for different devices showing conductive ($\pm1$, $\mp1$) states (no contact resistance correction). Conductance, $G$, is plotted as a function of the displacement field divided by the magnetic field, since the displacement value for the transitions scales roughly linearly with magnetic field. Conductance at $v_{tot}= 0$ (black lines) is compared to the conductance at $v_{tot}= -1$ (red lines) and $v_{tot}= +1$ (yellow lines). The observed states are given by the filling factor configurations in the top left corner. Cartoons depict outline of sample geometry, with edge lengths given in units of microns. Data for Figures 1-3 of the main text come from sample W. Fractional edge state measurements were partially performed in sample G. Note that the conductance of the $v_{tot}= +1$ states (yellow lines) is consistently smaller than the $v_{tot}= -1$ states (red lines) due to asymmetry in the contact resistances for negative and positive $v_{tot}$.

4-probe resistance measurements of helical edge states

If backscattering is possible between the helical edge states and spatially homogenous, then we expect that the edge resistance will scale linearly with the edge length. In this case, it follows straightforwardly that the 4-probe resistance measurement will depend on the edge lengths as $R_{4probe} \propto L_{SD}L_V / \sum L_i$, where $L_{SD}$ and $L_V$ are the edge lengths between the source-drain electrodes and voltage probes, respectively, and $\sum L_i$ is the sum of the edge lengths between contacts. In this situation of diffusive edge conductance, we would expect that the $R_{xx}$ measurement discussed in the main text would be *15.2* times greater than the $R_{NL}$ measurement, which is very close to the measured ratio at *B=1.5*T. But, as the magnetic field increases, the measurements converge, indicating a length-independent edge resistance.

In the absence of backscattering, a pair of helical edge states will act a ballistic 1d wire running along the edge of the sample. As is typical for 1d conductors, invasive contacts can interrupt the edge state by causing equilibration between the forward and backward moving modes. In this case, each edge segment between contacts will have a length-independent resistance of $h/e^2$. In a device with 4 contacts, a longitudinal resistance measurement (such as the $R_{NL}$ and $R_{xx}$ configuration discussed in the main text) will give a value of $h/4e^2$ since ¼ of the current flows through the quantum resistor between the two voltage probes. This is the value that the $R_{xx}$ and $R_{NL}$ measurements converge to with increasing magnetic field (Figure 3E of the main text).

Low magnetic field measurements

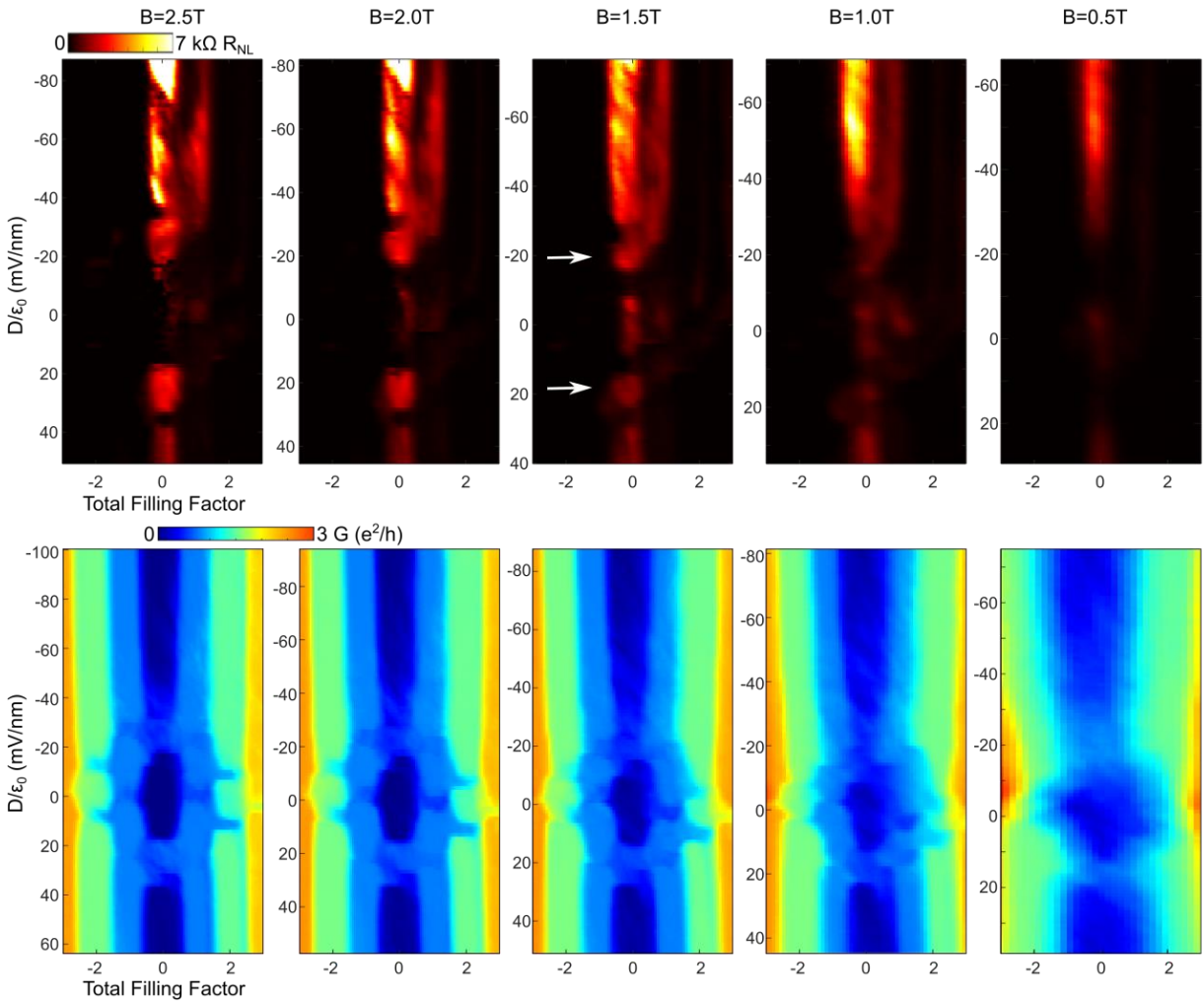

**Fig. S10**

Helical edge states onset at low magnetic fields. Data is for sample W presented in Figures 1-3 of the main text. Colorplots show the low magnetic field development of the nonlocal resistance (top) and the 2-probe conductance (bottom), as a function of displacement field and total filling factor. At $B = 1.5$ T, distinct nonlocal features can be seen ($\pm1, \mp1$). At the same magnetic field, clearly defined plateaus in conductance can be seen originating from the ($\pm1, \pm1$), ($0, \pm1$), ($\pm1, 0$) states. The data indicates well developed broken symmetry states and helical edge states at this low magnetic field.

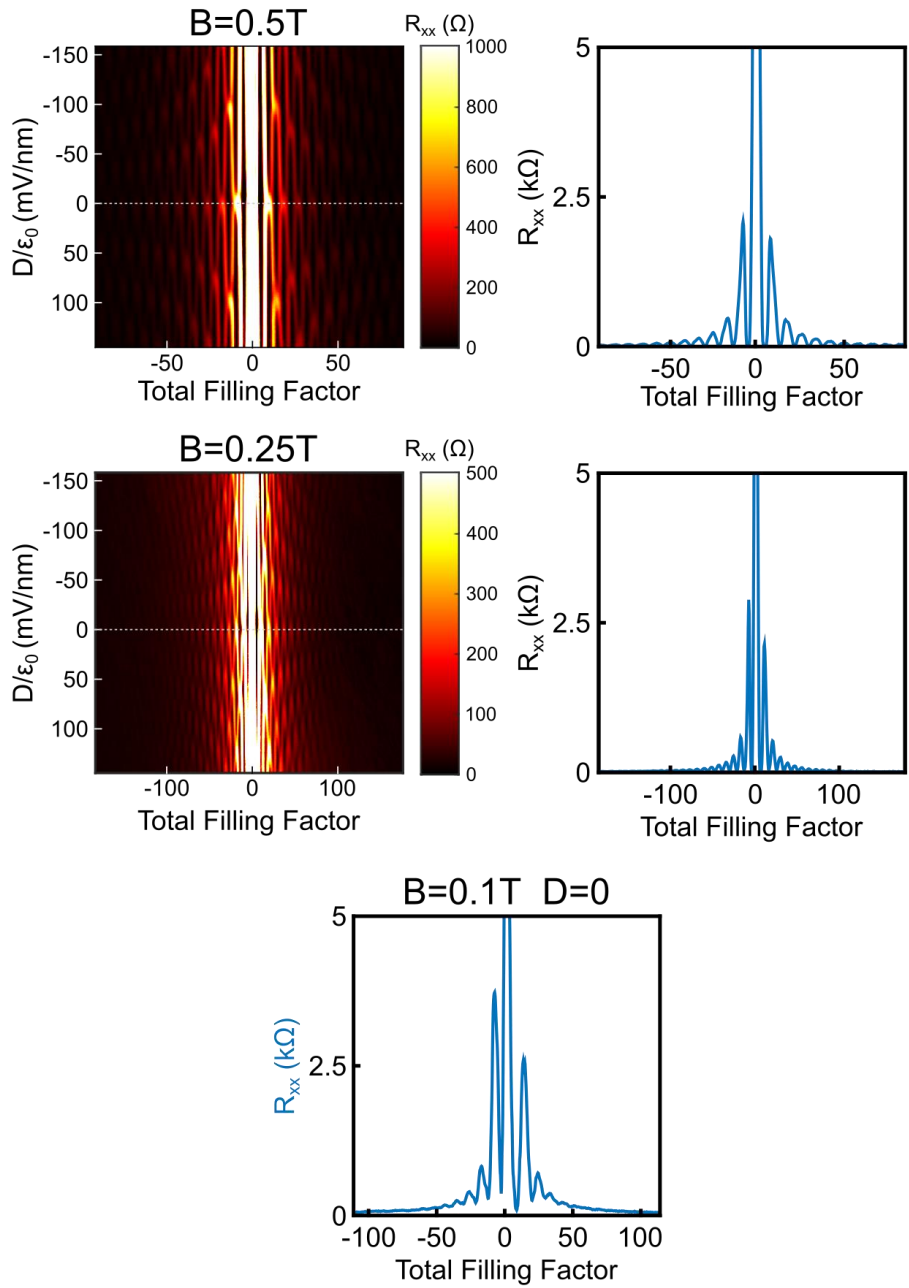

**Fig. S11**

Magnetoresistance at low magnetic fields consisting of 4-probe longitudinal resistance measurements for device W discussed in Figures 1, 2 & 3 of the main text. $R_{xx}$ peaks show Landau level crossing structure characteristic of twisted bilayer graphene devices (*26*). The well-developed $R_{xx}$ minima at such low magnetic fields highlight the high quality of this device.

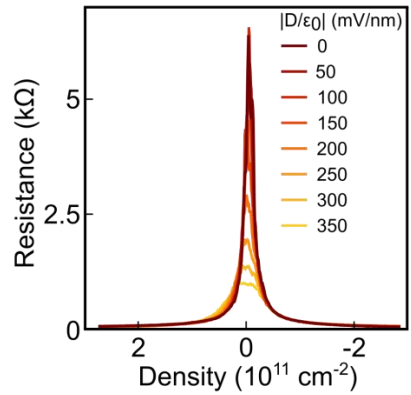

**Fig. S12**

At zero magnetic field, interlayer displacement field causes the charge-neutrality point resistance to decrease. 4-probe resistance at zero magnetic field as a function of total density and displacement field. The resistance at the charge neutrality point steadily decreases with displacement field (*26*). This behavior is in contrast to AB-bilayer graphene, where the effect of a displacement field is to open up a bandgap at the Dirac point (*34*), leading to a diverging resistance (*35*). Measurement is performed at *0.3* K.

Contact resistances and layer-selective contacts

All 2-probe measurements include the effects of contact resistance originating from the metal electrodes, the metal-graphene interface, and the graphene leading up to the primary device region. This is often seen as a suppression of the conductance of quantum Hall plateaus. We find that this effect can be corrected by subtracting a single contact resistance value for negative $v_{tot}$, and a different value for positive $v_{tot}$. As an example, the dataset in Figure 2D has a resistance of *0.6* and *2.0* kΩ subtracted from the negative and positive $v_{tot}$ parts of the dataset, respectively. This value of the contact resistance was extracted by fitting the value of the $v_{tot}=\pm 1$ to *1 $e^2$/h* and results in all the other conductance plateaus to line up with a quantized conductance value.

One cause of the asymmetry in the contact resistance between the negative and positive values of $v_{tot}$ comes from the formation of pn junctions in the graphene which lead to extra resistance. These arise because of changes in doping in the graphene between the dual gated region and the contacts. For example, Cr:Au electrodes tend to locally p-dope graphene near the contact, which will naturally cause a pn junction to arise when measuring n-doped graphene. Note that n-doping corresponds to positive $v_{tot}$ and p-doping to negative $v_{tot}$, since $v_{tot}$ is proportional to the number of electrons in the graphene measured relative to the charge neutrality point. The formation of pn junctions can be controlled by using extra gates to control the sign of the charge density outside of the primary device region under study. The effect of this can be seen in Figure S13, which shows the measurement of conductance plateaus in a twisted bilayer graphene device for p-doped contacts and n-doped contacts. In the case of p-doped contacts, there is no pn junction formed when measuring negative $v_{tot}$ and the plateaus have a very small contact resistance of order *100* Ω. By contrast, the positive $v_{tot}$ side has a very strongly suppressed conductance with the p-doped contacts. The map can be inverted by switching to n-doped contacts. To accomplish this contact doping control we use a device structure that has extra local topgates and a global Si backgate which dopes the twisted bilayer graphene all the way up to the metal electrodes. This allows us to tune the doping of the bilayer outside of the main region.

The formation of pn junctions at the contact interface presents a fundamental issue when measuring an electron-hole bilayer edge state. Because the state is made up of both p-doped and n-doped graphene, there will always be a pn junction formed when the contacts are of only one doping type. In fact, we observe that this effect can nearly shut off current injection into one of the layers, since the pn junction necessarily passes through zero density, which is insulating at high magnetic fields (*31*). A measurement of this effect is presented in Figure S14. Often, we observe that the conductance of the helical *(±1, ∓1)* states will be close to the conductance of the $v_{tot}$ = -1 plateau, as conductance is limited to only one of the layers. Using graphite contacts (as for sample W discussed in the main text), somewhat mitigates this problem since graphite has a similar work function to graphene and hence has less contact doping effects.

By having multiple gate-tunable contacts, it should be possible to have simultaneous independent contacts to both the top and bottom layers by having simultaneously contacts which are p-doped and n-doped. Figure S15 shows a test of this idea using a 4-probe device with independent topgates on each of its contacts to locally control the doping. As expected, the conductance of the *(-1,0)* state is maximized for p-doped contacts, and likewise the *(0,+1)* state conductance is maximized for n-doped contacts. By contrast, the *(-1,+1)* state conductance is maximized when there are two pairs of both n-doped and p-doped contacts. This is another confirmation of the electron-hole bilayer nature of the helical *(±1, ∓1)* states.

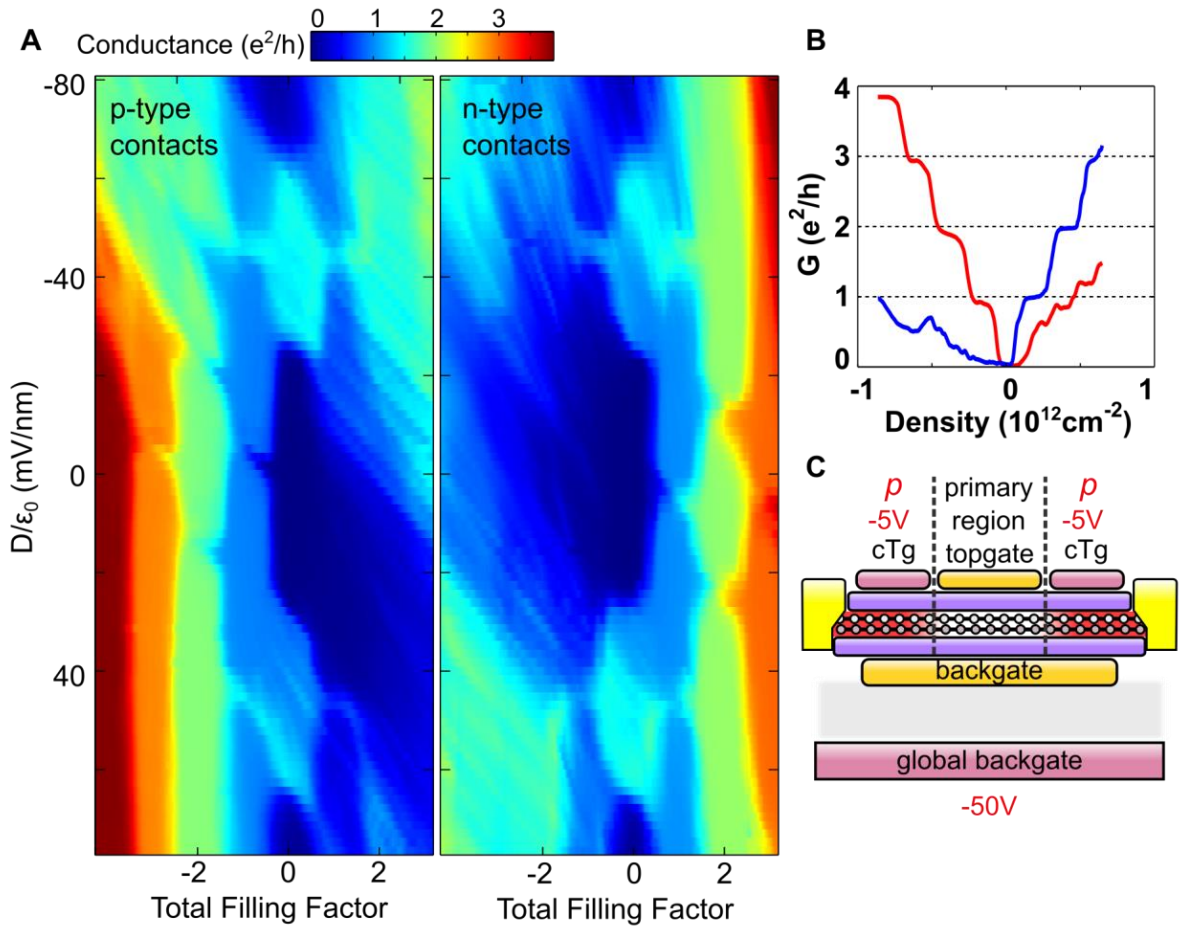

**Fig. S13**

Gate-tunable contacts can switch from making good contact to either negative $v_{tot}$ states or positive $v_{tot}$ states (Sample B). (A) Conductance maps for a device with gate tunable contacts. P-doped contacts result in clear measurements of the p-side of the data (negative filling factor) with strong suppression of the conductance for the n-side (positive filling factor). The converse is true for n-doped contacts. In both measurements the contact resistance in the well-measured plateaus is less than *100 Ω*. (B) Conductance plateaus for p-doped (red) and n-doped (blue) contacts. (C) Cross-section cartoon of device. The contact topgates (cTg) and the global backgate control the doping of the twisted bilayer graphene between the primary region of the device and the metal electrodes.

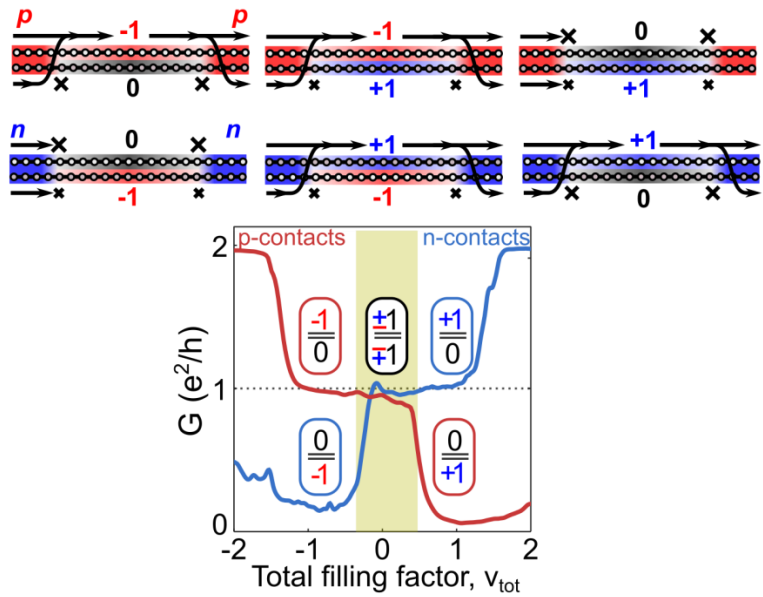

**Fig. S14**

Gate tunable contacts can selective inject current into only one layer (Sample B). Top – Cartoons show current flow paths depending on the filling factor of the layers and whether contacts are p-doped or n-doped. Current flow is blocked at pn junctions because the zero density state is insulating. Bottom – Conductance near the *(±1, ∓1)* states for p-dope and n-dope contacts. In the *(±1, ∓1)* states, the contacts can only inject current effectively into one layer.

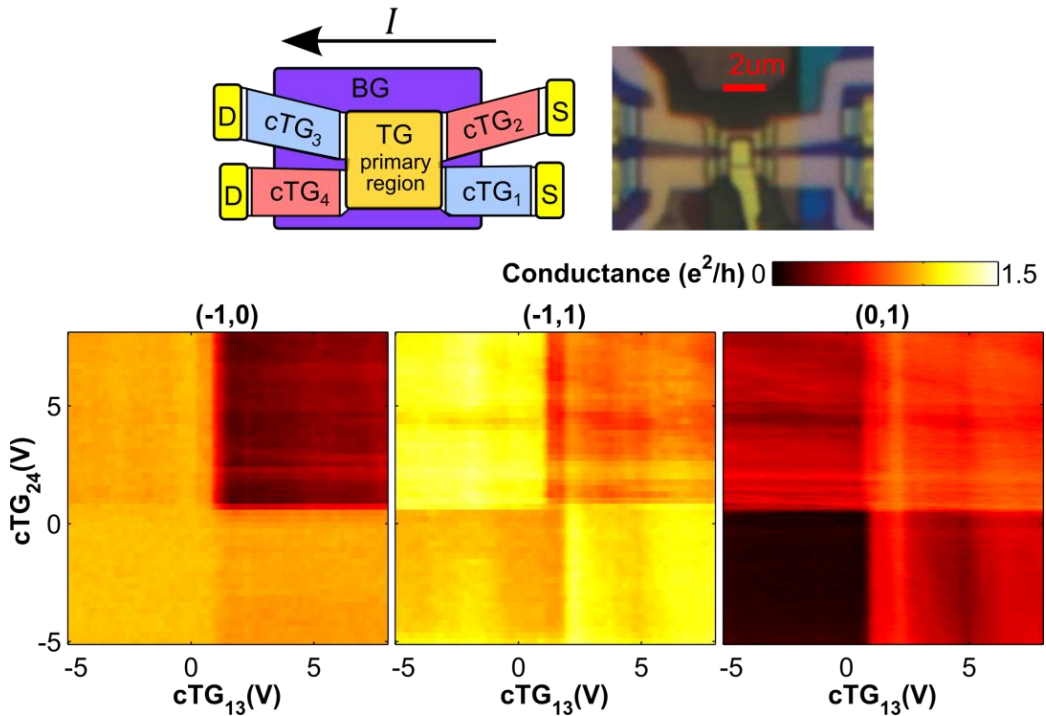

**Fig. S15**

Simultaneous p- and n-type contacts give the best measurement configuration (Sample B). Top – Cartoon schematic and optical image of device with four independent gate-tunable contacts labeled cTG$_{1,2,3,4}$. Bottom – Conductance maps in *(-1,0)*, *(-1,1)* and *(0,1)* states as a function of the contact topgate values. 2-probe conductance is measured between the source and drain electrodes (S and D in the top cartoon). cTG$_1$ and cTG$_3$ are swept together (cTG$_{13}$) and likewise for cTG$_2$ and cTG$_4$ (cTG$_{24}$). For the *(-1,0)* state, the highest conductance is for p-type contacts (negative cTG). Similarly, the *(0,1)* state has highest conductance for n-type contacts (positive cTG). By contrast, the *(-1,1)* state measurement has the highest conductance when there is simultaneous both p-type and n-type contacts (cTG$_{13}$ and cTG$_{24}$ opposite sign). This is further evidence that the *(-1,1)* state is made up of both p-type and n-type states.